\documentclass[]{aastex631}
\usepackage{CJK}
\shorttitle{AASTeX v6.31 Sample article}
\shortauthors{Chen et al.}
\graphicspath{{./}{figures/}}
\usepackage{soul}
\usepackage{color,xcolor,tcolorbox}
\usepackage{amsmath}
\usepackage{boondox-cal}
\usepackage{booktabs,multirow}

\begin{document}
\begin{CJK*}{UTF8}{gbsn}

\title{Coronal Mass Ejections Deflected by Newly Emerging Flux: A Combined Analytic and Numerical Study}

\author[0000-0002-8077-094X]{Yuhao Chen}
\affiliation{School of Earth and Space Sciences, Peking University, Beijing 100781, People’s Republic of China}
\affiliation{State Key Laboratory of Solar Activity and Space Weather, National Space Science Center, Chinese Academy of Sciences, Beijing 100190, People’s Republic of China}

\author[0000-0002-9258-4490]{Chengcai Shen}
\affiliation{Harvard-Smithsonian Center for Astrophysics, 60 Garden Street, Cambridge, MA 02138, USA}

\author[0000-0001-9650-1536]{Zhixing Mei}
\affiliation{Yunnan Observatories, Chinese Academy of Sciences, P.O. Box 110, Kunming, Yunnan 650216, People's Republic of China}
\affiliation{Yunnan Key Laboratory of Solar Physics and Space Science, Kunming, Yunnan 650216, People's Republic of China}

\author[0000-0002-5983-104X]{Jing Ye}
\affiliation{Yunnan Observatories, Chinese Academy of Sciences, P.O. Box 110, Kunming, Yunnan 650216, People's Republic of China}
\affiliation{Yunnan Key Laboratory of Solar Physics and Space Science, Kunming, Yunnan 650216, People's Republic of China}

\author[0000-0001-9828-1549]{Jialiang Hu}
\affiliation{State Key Laboratory of Solar Activity and Space Weather, National Astronomical observatories, Chinese Academy of Sciences, 100101, Beiing, China}
\affiliation{School of Astronomy and Space Science, University of Chinese Academy of Sciences, 100049, Beijing, China}

\author[0000-0003-0880-9616]{Zehao Tang}
\affiliation{Yunnan Observatories, Chinese Academy of Sciences, P.O. Box 110, Kunming, Yunnan 650216, People's Republic of China}
\affiliation{Yunnan Key Laboratory of Solar Physics and Space Science, Kunming, Yunnan 650216, People's Republic of China}

\author[0000-0002-1264-6971]{Guanchong Cheng}
\affiliation{Yunnan Observatories, Chinese Academy of Sciences, P.O. Box 110, Kunming, Yunnan 650216, People's Republic of China}
\affiliation{Yunnan Key Laboratory of Solar Physics and Space Science, Kunming, Yunnan 650216, People's Republic of China}

\author[0000-0003-2465-5231]{Shanshan Xu}
\affiliation{Key Laboratory of Particle Astrophyics and Experimental Physics Division and Computing Center,
Institute of High Energy Physics, Chinese Academy of Sciences, 100049 Beijing, China}

\author[0000-0002-2388-7068]{Abdullah Zafar}
\affiliation{Yunnan Observatories, Chinese Academy of Sciences, P.O. Box 110, Kunming, Yunnan 650216, People's Republic of China}

\author[0009-0001-0703-2000]{Yujia Song}
\affiliation{School of Astronomy and Space Science, University of Chinese Academy of Sciences, 100049, Beijing, China}
\affiliation{Key Laboratory for Computational Astrophysics, National Astronomical Observatories, Chinese Academy of Sciences, Datun Road A20, Beijing 100012,China}

\author[0000-0002-3326-5860]{Jun Lin}
\affiliation{Yunnan Observatories, Chinese Academy of Sciences, P.O. Box 110, Kunming, Yunnan 650216, People's Republic of China}
\affiliation{Yunnan Key Laboratory of Solar Physics and Space Science, Kunming, Yunnan 650216, People's Republic of China}
\affiliation{University of Chinese Academy of Sciences, Beijing 100049, People's Republic of China}

\correspondingauthor{Zhixing Mei, Jing Ye}
\email{meizhixing@ynao.ac.cn, yj@ynao.ac.cn}




\begin{abstract}
Newly emerging flux (NEF) has been widely studied as a trigger of solar filament eruptions, but its influence on the subsequent dynamics remains poorly explored. Because NEF typically emerges adjacent to filaments, it imposes magnetic asymmetry that can drive non-radial eruptions and complicate space-weather forecasting. We bridge analytic catastrophe theory with 2D resistive MHD simulations: analytic solutions provide magnetic configurations containing a flux rope at the loss-of-equilibrium point, which are then used as initial conditions for simulations to examine the following dynamics. We find that NEF governs the kinematics of filament eruptions in two ways. First, by reshaping coronal stability, NEF can create or eliminate a higher equilibrium in corona, thereby producing failed eruptions or CMEs. In the transitional situation where a metastable equilibrium appears, the rising filament decelerates and stalls before re-accelerating into a CME, consistent with observed two-step eruptions. Second, by breaking symmetry, NEF deflects eruptions away from the radial direction: depending on its polarity, it acts as a repulsor or an attractor on eruptive filaments, and the deflection magnitude increases with the degree of asymmetry. Our theory yields two characteristic angles that predict the deflection directions of CMEs and failed eruptions, and simulations closely aligns with these predictors. These results highlight the NEF not only as a trigger but also as a key factor that governs both the acceleration and deflection of eruptions during their propagation in the low corona.
\end{abstract}



\section{Introduction} \label{sec:intro}

Solar filaments appear as dark features on the solar disk, containing dense and cool plasma. Under certain conditions, these initially stable filaments may lose their equilibrium and enter a dynamic eruptive phase. Such eruptions are frequently associated with coronal mass ejections (CMEs) \citep{2003ApJ...586..562G,2013AdSpR..51.1967S,2015ASSL..415..411W}. Once CMEs reach Earth, their magnetic fields can reconnect with and/or compress the magnetosphere, driving geomagnetic storms and affecting the power grids. Additionally, CME-driven shocks accelerate high-energy particles that disrupt satellite operations and pose serious radiation hazards to astronauts. Therefore, understanding the CME initiation, associated high-energy particles production, and subsequent propagation is essential for reliable space-weather forecasting \citep{2015ASSL..415..433L}.

Two fundamental questions are central to understanding filament eruptions and their space weather effects: how the eruption is triggered, and how the eruption subsequently propagates through the corona and interplanetary space. The former addresses the initiation of eruption, while the latter determines whether the ejected plasma will ultimately affect the Earth. Together, they constitute a complete picture of solar eruptions. Substantial progress has been made in understanding the triggering mechanism. As reviewed by \citet{2000JGR...10523153F}, models powering CMEs are commonly grouped into three categories, including ideal MHD models such as catastrophe \citep{1991ApJ...373..294F}, torus instability \citep{2006PhRvL..96y5002K}, and kink instability \citep{2004A&A...413L..27T}; resistive models involving reconnection such as the breakout \citep{1999ApJ...510..485A} and tether-cutting \citep{2001ApJ...552..833M}; and hybrid models exemplified by the Lin--Forbes model \citep{2000JGR...105.2375L}. 

Furthermore, the propagation of CMEs has been extensively investigated over the past decades \citep[see reviews by][]{2011LRSP....8....1C, 2017SSRv..212.1159M}. These studies have shown that CMEs and their interplanetary counterparts (ICMEs) can undergo various physical processes during their propagation from the corona to the heliosphere, including various ways such as rotation, deformation, deflection, and deceleration. \citet{2017SoPh..292..118S} demonstrated that Lorentz forces dominate the CME propagation in the low corona, while the aerodynamic drag caused by the interaction with the solar wind becomes increasingly important at greater heliocentric distances \citep[e.g., see][]{2004SoPh..221..135C, 2012GeoRL..3919107S}. In this work, we mainly focus on the early propagation of CMEs evolved from eruptive filaments in the low corona, where the complex coronal magnetic field strongly constrains the CME dynamics, and a quantitative analysis of their erupting trajectory remains limited. A statistical study by \citet{2015SoPh..290.1703M} further shows that the propagation of eruptive filaments generally exhibits complex dynamics. These complexities primarily arise from two aspects.

First, the eruptive filament often deviates from the radial direction during its outward propagation. Such non-radial motion is typically caused by the nearby magnetic structures in the low corona. \cite{2023FrASS..1060432C} reviewed recent observational and numerical studies of CME deflections and classified these background structures into two families: repulsors and attractors. Repulsors, acting like magnetic walls, tend to deflect the erupting filament away from them. Similar instances can be found in coronal holes with open magnetic field lines \citep{2006AdSpR..38..461C,2009JGRA..114.0A22G,2020ApJ...899....6M,2020ApJ...896...53S,2020AdSpR..65.1654C} and active regions with strong magnetic fields \citep{2015NatCo...6.7135M,2015ApJ...805..168K,2015ApJ...814...80W}. In contrast, attractors behave more like magnetic wells, guiding the eruption toward them. These typically include regions with relatively low magnetic energy, such as the heliospheric current sheet \citep{2011SoPh..269..389S,2011SoPh..271..111G,2015SoPh..290.3343L,2020JGRA..12527530W}, pseudostreamers \citep{2013ApJ...773..162B,2013ApJ...764...87L,2020JGRA..12527530W}, and helmet streamers \citep{2012ApJ...749...12Y,2018ApJ...862...86Y}. Beyond deflection, recent observations suggest that pseudostreamer configurations may play an important role in ``double-bang firecracker'' events involving filament splitting \citep{2025SCPMA..6879611Z}. Moreover, as the CME further enters interplanetary space, its trajectory can also be affected by the background solar wind \citep{2004SoPh..222..329W} and interactions with other CMEs \citep{2012ApJ...759...68L}. These deflection processes critically affect the space weather impact of CMEs. For example, \cite{paouris2025cme} showed that the CME occurring on April 21, 2023, although not strongly Earth-directed initially, was deflected toward the solar equator by a southern coronal hole, resulting in an unexpectedly severe geomagnetic storm.

Second, eruptive filaments exhibit distinct dynamical modes following their loss of equilibrium. In the simplest and most common mode, termed a ``one-step eruption,'' the filament continuously accelerates after losing equilibrium and smoothly evolves into a CME. In a more intricate mode, known as a ``two-step eruption,'' the filament initially ascends rapidly but subsequently experiences significant deceleration or even stagnation near a new quasi-equilibrium height, where it may stay for several hours before undergoing secondary acceleration and eventually developing into a CME \citep{2014SoPh..289.4545B,2016ApJ...821...85G,2017SoPh..292...81C,2018MNRAS.475.1646F}. In the third scenario, clearly distinct from the previous two CME-producing or successful eruption cases, eruptive filaments rise to a new equilibrium height without further ascent, termed a ``failed eruption'' \citep{2003ApJ...595L.135J,2011SoPh..272..301K,2013ApJ...778...70C,2022ApJ...933..148C}. These dynamical modes have direct implications for CME forecasting. Whether the eruption is successful or failed determines whether a CME will occur at all, while the difference between one-step and two-step eruptions affects the onset timing of CME, which influences the predicted arrival time at Earth.

In particular, newly emerging flux (NEF) commonly appears alongside filament channels and ubiquitously introduces asymmetries in photospheric flux distributions and coronal connectivity, making it a natural contributor to the complex dynamics of filament eruptions \citep{2015SoPh..290.1703M}. Motivated by this connection, we focus on how NEF affects the propagation of eruptive filaments in this paper. 

Solar active regions form through the magnetic flux emergence from the solar interior, and therefore the NEF has been widely studied as a key trigger for filament eruptions (see reviews by \citealt{2011LRSP....8....1C} and \citealt{2014LRSP...11....3C}). A widely accepted mechanism is that filament eruptions are primarily triggered when NEF is more likely to reconnect with the coronal background field, whereas the filament typically remains stable when the NEF is reconnection-unfavorable \citep{1995JGR...100.3355F,2000ApJ...545..524C}. However, recent analytic and numerical studies indicate that even the reconnection-unfavorable NEF can occasionally lead to filament eruption, revealing the unresolved condition for NEF-driven eruptions \citep{2001JGR...10625053L,2022ApJ...933..148C,2024ApJ...977L..26C}. Recent observations further reveal that eruption-triggering flux does not always arise from photospheric emergence, but may be fed by repeated mini-filament eruptions beneath a giant filament \citep{Chen_2025}.

While our understanding of NEF-triggered instabilities in filaments has improved, the impact of NEF on their subsequent propagation remains largely unexplored. Based on an ideal MHD model, \cite{2024ApJ...977L..26C} analytically studied the quasi-static evolution of a filament in response to the NEF, and they noted that the NEF not only acts as a trigger of eruption, but also reshapes the coronal field in two aspects. On one hand, NEF introduces asymmetries into the coronal configuration and may cause substantial deflections. On the other hand, NEF modifies the equilibrium positions of the filament in the corona and may further influence its dynamical mode. These findings imply that NEF plays a crucial role in determining the propagation process of eruptive filaments—a mechanism that has yet to be systematically investigated.

To address the role of NEF, we perform two-dimensional resistive Magnetohydrodynamics (MHD) simulations based on \cite{2024ApJ...977L..26C} to systematically explore how the location and polarity of NEF influence eruptive filaments' propagation trajectories and dynamical modes. The paper is organized as follows: Section \ref{sec:model} describes the analytic and numerical methods. In Section \ref{sec:result} we present the results. In Section \ref{sec:discussions}, we provide discussions. Finally, we summarize this work in Section \ref{sec:conclusion}.

\section{Methods} \label{sec:model}

We adopt a hybrid approach that combines analytic model and numerical simulations. These two methods offer complementary advantages and are used sequentially to describe different phases of the eruption.

Analytic models are particularly effective in describing the quasi-static evolution of the system and in determining the critical conditions under which a filament loses equilibrium or the catastrophe occurs. They provide precise mathematical expressions that capture the system's behavior across the full parameter space \citep[e.g.,][]{2024ApJ...977L..26C}. However, once the filament erupts, a long current sheet forms behind it and magnetic reconnection becomes essential to the evolution \citep{2000JGR...105.2375L}. This highly dynamic, nonlinear process is difficult to treat analytically due to its inherent mathematical complexity.

Additionally, numerical simulations are well suited to modeling the magnetic effect during eruptions \citep{2012MNRAS.425.2824M,2019MNRAS.482..588Y,2023ApJ...955...88Y}. However, they are less effective in the pre-eruption phase for two main reasons. First, the energy buildup driven by slow photospheric motions typically occurs over timescales of tens of hours to days or even weeks \citep{2000JGR...105.2375L,2012ApJ...748...77S,2018SSRv..214...46G}, making direct simulations computationally expensive and prone to cumulative numerical errors. Second, simulations are limited to case-by-case studies and do not offer a global view of the system's behavior across the full parameter space.

Therefore, we first use the analytic model developed by \citet{2024ApJ...977L..26C} to identify the critical state, where the NEF drives the system to a catastrophe. We then use these critical states as initial conditions in our two-dimensional resistive MHD simulations, which calculate the subsequent dynamic evolution. This strategy bridges the quasi-static and dynamic phases of a filament eruption and minimizes numerical errors.

\subsection{Analytic Model} \label{sec:model_am}

\begin{figure}[ht]
\begin{center}
\includegraphics[width=0.4\textwidth]{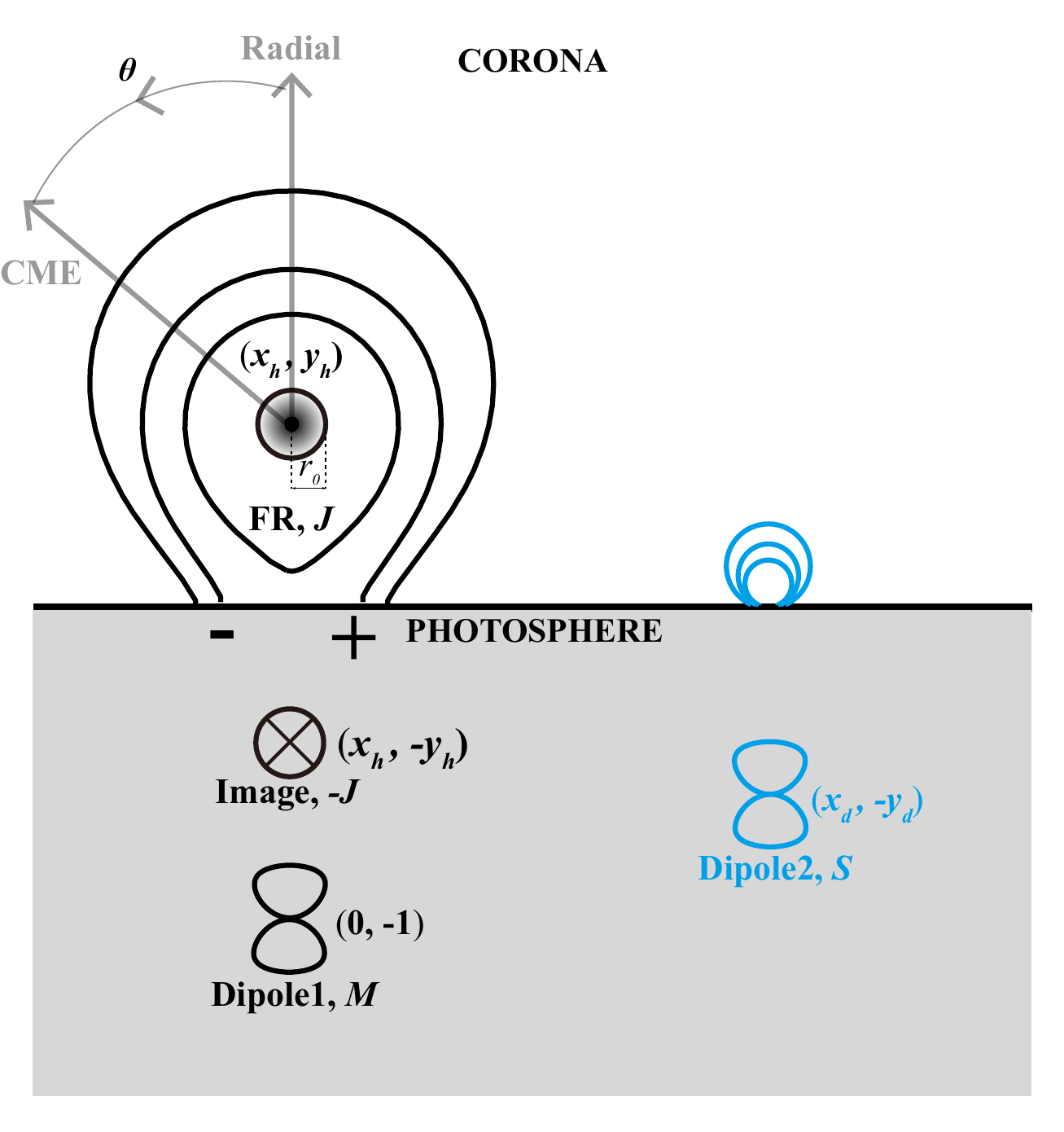}
\end{center}
\caption{Schematic of the magnetic configuration produced by four components: the flux rope, the image of the flux rope, dipole 1 to generate the unchanged background field, and dipole 2 to model the NEF. The deflection angle $\theta$ is the angle between the CME direction and the solar radial direction. By convention, $-90^{\circ}<\theta<0$ when the CME deviates clockwise from the radial direction, and $0<\theta<90^{\circ}$ when it deviates counterclockwise. Since we only consider outward propagation ($v_y>0$), $\theta$ is restricted to $(-90^{\circ},\,90^{\circ})$, eliminating any $180^{\circ}$ ambiguity. This figure is reproduced from \cite{2024ApJ...977L..26C}
\label{fig:config}}
\end{figure}

Figure \ref{fig:config} presents the magnetic configuration adopted from \cite{2024ApJ...977L..26C}. The background field is generated by the unchanged dipole 1 beneath the photosphere. The flux rope carrying current represents the filament in the corona, which induces an image current beneath the photosphere. The NEF is produced by the dipole 2. By gradually increasing the strength of dipole 2, we simulate the slow flux emergence and the resulting equilibrium evolution of the flux rope. 

The positions of the flux rope, dipole 1, and dipole 2 are defined at ($x_h, y_h$), (0, -1), and ($x_d, -y_d$), while their strengths are set by parameters $J$, $M$, and $S$, respectively. All these quantities are dimensionless, normalized by the characteristic length $d = 4 \times 10^{7}$~m and the magnetic field $B_{0} = 2.8$~G, related by $B_0=2I_0/cd$, where $d$ denotes the characteristic height of the filament, $B_{0}$ represents the characteristic strength of the coronal magnetic field, $I_{0}$ is the electric current inside the filament, and $c$ is the light speed. The photosphere is located at $y = 0$. The solar radial direction is taken along the $y$- axis, and the deflection angle $\theta$ is defined between this radial direction and the line from the initial flux-rope position to the CME trajectory. We take $\theta<0$ when the CME deviates clockwise from the radial direction, and $\theta>0$ when it deviates counterclockwise. The flux rope radius is $r_0$, with its initial value $r_{00}$ defined for $J = 1$. We set $r_{00} = 0.05$, corresponding to $2 \times 10^{6}$~m, which is comparable to the typical radius of observed filaments. This value is larger than the $r_{00} = 0.01$ adopted by \cite{2024ApJ...977L..26C}, ensuring sufficient grid resolution inside the flux rope in following simulations. 

\begin{figure}[ht]
\begin{center}
\includegraphics[width=0.95\textwidth]{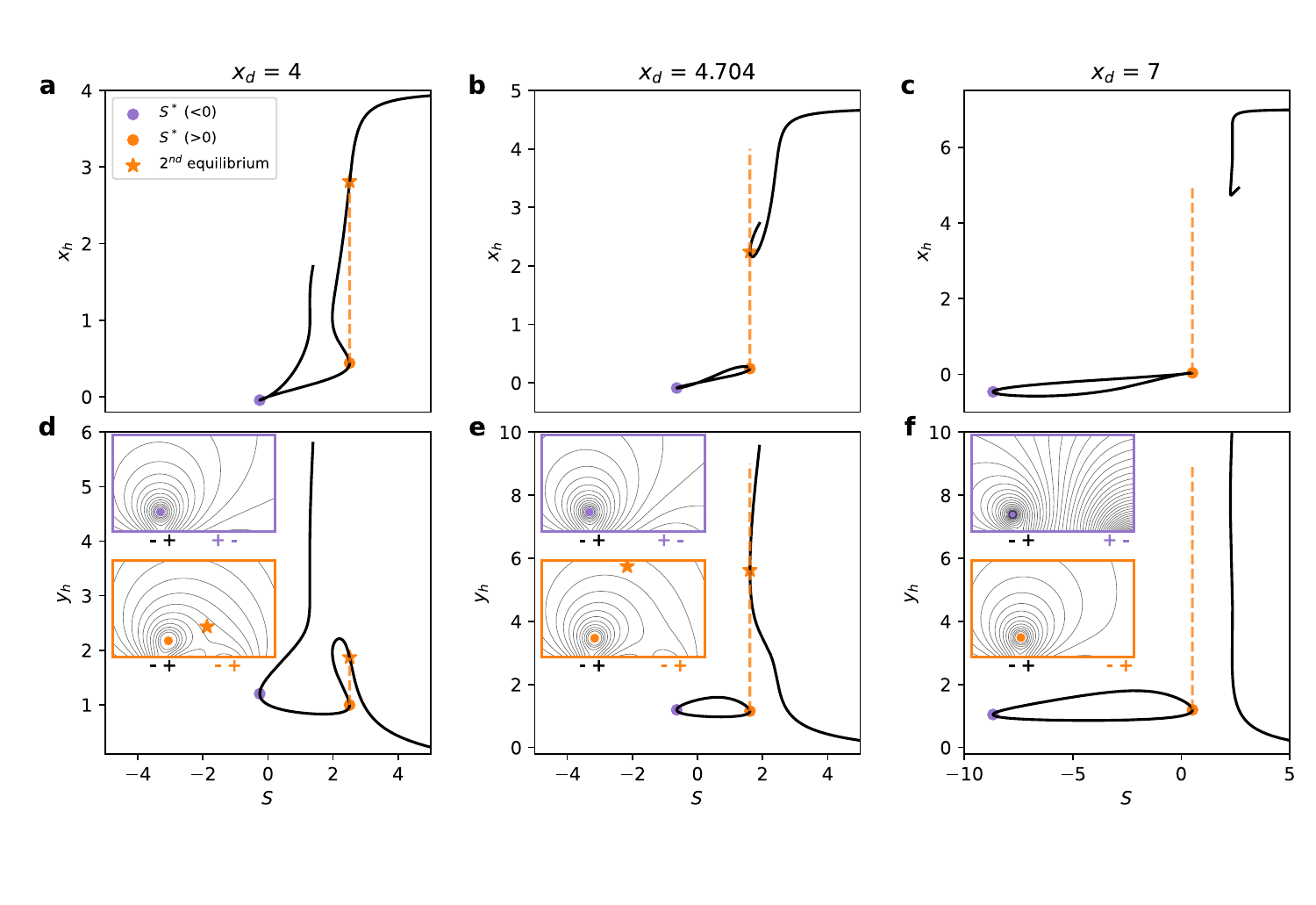}
\end{center}
\caption{Equilibrium evolution as the NEF strength $S$ increases quasi-statically given $y_d=2.5$ and $r_{00}=0.05$. From left to right, the columns show cases with the NEF fixed at $x_d=4$, $4.704$, and $7$. The top row plots the horizontal equilibrium position $x_h(S)$; the bottom row plots the height $y_h(S)$. Black curves are equilibrium branches.
Dots denote the critical points where catastrophe occurs: purple for $S^*<0$ and orange for $S^*>0$. Orange stars mark a second equilibrium that may be reached after the loss of the original equilibrium at the same $S^*$; the orange dashed line indicates $S=S^*$.
Insets in bottom panels illustrate the magnetic configuration at the corresponding critical states; border colors match that for critical points.
These critical configurations are used as the initial conditions for following MHD simulations.
\label{fig:curve}}
\end{figure}

Following \citet{2001JGR...10625053L} and \citet{2024ApJ...977L..26C}, the free magnetic energy stored in the system, $E_h$, is given by:
\begin{align}
E_h = \left( \frac{I_0}{c} \right)^2 J^2 \left[ \ln \left( \frac{2 y_h J}{r_{00}} \right) + \frac{1}{2} \right]. \label{Eh}
\end{align} 
Here the free energy is the difference between the total energy in the system and the energy in the corresponding
potential system. The flux rope reaches equilibrium when the Lorentz force vanishes:
\begin{align}
\frac{\partial E_h }{\partial x_h}= 0, \label{Global eqx} \\ 
\frac{\partial E_h }{\partial y_h}= 0.
\label{Global eqy}
\end{align}
Equilibrium requires that the total Lorentz force on the flux rope vanishes. \cite{2001JGR...10625053L} suggested that the total equilibrium can be separated into two parts: a global equilibrium governed by the external field acting on the flux rope (Equations~\ref{Global eqx} and \ref{Global eqy}), and a local equilibrium that ensures the rope’s self Lorentz force is internally balanced (see \citealt{2024ApJ...977L..26C}). By combining global and local equilibrium conditions, we derive how the equilibrium position of the flux rope varies with $S$ (black curves in Figure \ref{fig:curve}). The critical points on these curves, marking the onset of catastrophe, are determined by
\begin{align}
\left( \frac{\partial^2 E_h }{\partial x_h^2} \right)\left( \frac{\partial^2 E_h }{\partial y_h^2} \right)-\left( \frac{\partial^2 E_h }{\partial x_h \partial y_h} \right)^2= 0, 
\label{critical point}
\end{align}
and are denoted as $(S^*, x_h^*)$ and $(S^*, y_h^*)$ (purple and orange dots in Figure~\ref{fig:curve}). Here, $x_h^*$ denotes the horizontal position of the flux rope at the critical point, and $y_h^*$ denotes its vertical position. Further details of the derivation can be found in \citet{2001JGR...10625053L} and \citet{2024ApJ...977L..26C}.

There are two equivalent ways to represent the emergence of new magnetic flux: (1) fixing the magnetic source depth $y_d$ while increasing its field strength $S$, or (2) fixing $S$ while moving the source upward by decreasing $y_d$. Both approaches can reproduce the gradual appearance of magnetic flux at the photosphere and have been used in previous analytic and numerical studies \citep[e.g., see][]{2001JGR...10625053L,2022ApJ...933..148C,2024ApJ...977L..26C}. In this work, we adopt the first approach, fixing $y_d$ and varying $S$, which provides a convenient way to represent the status without NEF at the beginning by setting $S=0$. Increasing $S$ from zero at a fixed depth satisfies this physical requirement and is straightforward to implement in analytic models. Since the dynamics of magnetic emergence below the photosphere are not fully understood, this approach represents a reasonable approximation of the magnetic flux emergence process.

Furthermore, the choice of different emergence approaches only affects the quasi-static evolution of the whole magnetic configuration but not the critical state at which the eruption is triggered. Mathematically, the critical point represents a specific magnetic configuration which can be described by a closed set of equations (Eqs. \ref{Global eqx}-\ref{critical point} and \ref{forzenflux_eq}) that relate the parameters $(x_d, y_d, S, x_h, y_h, J)$; fixing any two of these quantities uniquely specifies the remaining four. For example, if the system evolves via approach (1), increasing $S$ at a fixed position $(x_d,y_d^*)$, the critical parameters are $(S^*,x_h^*,y_h^*,J^*)$. Conversely, if we consider approach (2) at a fixed $(x_d,S^*)$, the equilibrium equations yield the same critical state when $y_d$ reaches to $y_d^*$, i.e., $(y_d^*,x_h^*,y_h^*,J^*)$. Hence, the choice of emergence approach only affects the quasi-static evolution but not the critical state at which the eruption is triggered. Therefore, the subsequent numerical simulations initiated from this critical point are representative of both emergence approaches.

In this study, we fix $y_d = 2.5$, and mainly study the effects by varying $x_d$ and $S$. We choose $y_d = 2.5$ because analytic solutions indicate that a higher equilibrium position exists after the catastrophe only if the value of $y_d$ is not too small (e.g., $\ge 2$) \citep{2024ApJ...977L..26C}; at the other extreme, for substantially larger depths the NEF exerts only a weak influence on the pre-existing flux rope and barely alters its equilibrium. Thus $y_d = 2.5$ samples the intermediate regime where the rich eruption behaviors including successful and failed eruptions can occur.

Additionally, we restrict our setup to scenarios with $7\ge x_d\ge2.5$. As discussed by \cite{2024ApJ...977L..26C}, a critical point always exists regardless of $x_d$, whereas for $S>0$, the occurrence of catastrophe depends on $x_d$. If the NEF emerges very close to the filament (e.g., $x_d<2$) and $S>0$, the system tends to become stable, preventing the catastrophe. In such cases, NEF is unfavorable for reconnecting with the background field and acts to constrain the flux rope (see simulation results in \citealt{2000ApJ...545..524C} and recent observations in \citealt{2025ApJ...987L..21Z}). On the other hand, the catastrophe can still occur for larger $x_d$, but as the NEF moves farther away, its influence on the pre-existing flux rope becomes so weak that its effect on the equilibrium evolution and eruption dynamics becomes negligible. In previous analytical models, \cite{2024ApJ...977L..26C} showed that for $S<0$ and $x_d>7$, the NEF would need to be more than an order of magnitude stronger than the background field to trigger an eruption. Therefore, we adopted $x_d = 7$ as the upper limit. Overall, the range $7\ge x_d\ge2.5$ sufficiently covers the most important regime to study eruptions of the flux rope. 

Figure~\ref{fig:curve} illustrates how the equilibrium position of the flux rope evolves with increasing $|S|$ for three representative horizontal positions of the NEF ($x_d=4$, $4.704$, and $7$). The top and bottom panels depict the equilibrium evolutions in both $x_h$- and $y_h$- directions, respectively. The subplots in the second column show magnetic configurations at critical states, where the filament is expected to loss equilibrium, with boundary colors matching their respective critical points. These critical magnetic configurations will be utilized as the initial condition in simulations to study the subsequent dynamic process. It is worth noting that orange stars indicate the possible secondary equilibrium positions after the catastrophe, as the flux rope may jump to an upper equilibrium state. Physical implications of the equilibrium curves are discussed in Section \ref{sec:result_analytic}.

\subsection{Numerical Approach}

To investigate the dynamic evolution of eruptive filaments, we follow the numerical framework in \cite{2022ApJ...933..148C}. The open-source NIRVANA code \citep{2004JCoPh.196..393Z,2005CoPhC.170..153Z,2008CoPhC.179..227Z}t is used to solve the full 2D resistive MHD equations:
\begin{eqnarray}
\partial_t \rho + \nabla\cdot(\rho{\bf{v}})&=&0,\\
\partial_t e+\nabla\cdot\left[(e+p+\frac{B^2}{2\mu}){\bf{v}}-\frac{1}{\mu}(\bf{v}\cdot \bf{B})\bf{B}\right]&=&\rho {\bf{g}}\cdot{\bf{v}}+\nabla\cdot\left[ \bf{v}\tau+\frac{\eta}{\mu}\bf{B}\times (\nabla\times\bf{B})\right]\label{energy},\quad\quad\\
\partial_t(\rho {\bf{v}})+\nabla\cdot\left[\rho{\bf{v}}{\bf{v}}+(p+\frac{B^2}{2\mu}){\bf{I}}-\frac{1}{\mu}\bf{BB}\right]&=&\nabla\cdot\tau+\rho\bf{g},\\
\partial_t\bf{B}&=&\nabla\times(\bf{v}\times \bf{B}-\eta\nabla\times\bf{B})\label{Induction},\\
\nabla\cdot\bf{B}&=&0\label{Divergence free},
\end{eqnarray}
where $\rho$, $\mathbf{v}$, $p$, $\mathbf{B}$, $e$, and $\mathbf{g}$ are plasma density, velocity, gas pressure, magnetic field, total energy density, and gravity. $\mu$ is the permeability. The plasma is fully ionized and obeys the ideal gas law with $\gamma=5/3$. $\tau=\nu_d [\nabla \textbf{v}+(\nabla\textbf{v})^T-2(\nabla\cdot\textbf{v})\textbf{I}/3]$ is the stress tensor. $\nu_d$ and $\nu_k$ represent the dynamic and kinematic viscosity, respectively, where
$\nu_d=\rho \nu_k$ and $\nu_k=10^{8}$~m$^{2}$~s$^{-1}$. We set the magnetic resistivity to $\eta=5\times10^8$~m$^2$~s$^{-1}$ for the whole simulation domain to facilitate the fast reconnection. Based on parameters for a typical CME, the characteristic Alfv\'en speed and length scale can be taken as 
$v_A=10^3$~km~s$^{-1}$ and $L_0 = 10^5$~km, 
respectively, giving a magnetic Reynolds number of $R_m = v_A L_0/\eta= 2 \times 10^5$. The simulations domain covers $[-4L_0,\,4L_0]\times[0,\,8L_0]$, which is substantially larger than the region of interest to minimize boundary effects. Gravity yields a stratified atmosphere with a photosphere ($0\le y\le h_p$), chromosphere ($h_p\le y\le h_c$), and corona ($y\ge h_c$), where $h_p=h_c=10^6$~m (see details in \citealp{2012MNRAS.425.2824M,2019MNRAS.490.2918X,2022MNRAS.509..406X}). The photosphere is isothermal with $T_p=4300$~K, and the coronal base temperature is $T_c=1.06\times10^6$~K. A base grid of $800\times800$ is used, with static refinement applied in the lower atmosphere and the region swept by the eruptive flux rope, yielding a finest cell size of $\delta=1.5625\times10^{-3}L_0$.

As discussed above, we performed a set of MHD simulations and each run is initialized from the critical magnetic configuration predicted by the analytic model (bottom panels of Figure \ref{fig:curve}). For a given $x_d$ and polarity, the critical parameters $(x_h^*,y_h^*,J^*,S^*)$ are obtained from Equation \ref{critical point}; together they uniquely specify the initial magnetic field (implementation details in \citealt{2022ApJ...933..148C}). We survey both negative-polarity ($S<0$) and positive-polarity ($S>0$), and horizontal locations $x_d$ in the range $2.5 \le x_d \le 7$, while fixing $y_d=2.5$. For each polarity, $x_d$ was first varied from 2.5 to 7.0 with an interval of 1.0, and additional runs with finer sampling between $x_d=4$ and $5$ were carried out for the $S>0$ cases to investigate the intermediate state around $x_d=4.704$. The critical parameters used as initial conditions are listed in Table \ref{tab:init}. The current within flux rope is prescribed with a smooth profile \citep{1990JGR....9511919F}, and the atmosphere is initialized in hydrostatic equilibrium with $\mathbf{v}=0$ at $t=0$.

\begin{deluxetable}{cDDDDD}
\tabletypesize{\footnotesize}
\tablewidth{0pt}
\tablecaption{Parameters describing the initial magnetic configuration in simulations.\label{tab:init}}
\tablehead{
\colhead{$x_d$} &
\twocolhead{$S^{*}$} &
\twocolhead{$x_h^{*}$} &
\twocolhead{$y_h^{*}$} &
\twocolhead{$J^{*}$}
}
\decimals
\startdata
\cutinhead{$S<0$}
2.5  &-0.063     &-0.016     &1.209    &0.985 \\
3.0  &-0.094     &-0.022     &1.206     &0.987 \\
4.0  &-0.253     &-0.045     &1.200     &0.994 \\
5.0  &-0.994     &-0.125     &1.187     &1.021 \\
6.0  &-3.713     &-0.306     &1.131     &1.096 \\
7.0  &-8.674     &-0.465     &1.044     &1.192 \\
\cutinhead{$S>0$}
2.5  & 4.376     & 0.656     &0.357     &0.613 \\
3.0  & 3.675     & 0.615     &0.583     &0.738 \\
4.0  & 2.509     & 0.440     &0.995     &0.858 \\
4.25 & 2.196     & 0.372     &1.070     &0.879 \\
4.50 & 1.873     & 0.300     &1.124     &0.899 \\
4.60 & 1.744     & 0.272     &1.140     &0.906 \\
4.70 & 1.617     & 0.245     &1.153     &0.913 \\
5.0  & 1.265     & 0.173     &1.178     &0.932 \\
6.0  & 0.656     & 0.064     &1.195     &0.961 \\
7.0  & 0.518     & 0.037     &1.195     &0.969 \\
\enddata
\tablecomments{%
Values are taken at the critical state. Units follow the normalization used in Section \ref{sec:model_am}.}
\end{deluxetable}

Outflow conditions are imposed on the left, right, and top boundaries by zero-gradient extrapolation of the primitive variables, allowing disturbances to exit the domain. At the bottom, line-tied boundary conditions \citep{1990JGR....9511919F,2011ApJ...737...14S} are implemented following \cite{2012MNRAS.425.2824M}: two high-density layers are introduced to anchor the magnetic field lines to the photosphere, with all quantities fixed in time.

It is worth noting that our MHD simulation does not address the quasi-static evolution of the background field in this work; instead, we start the simulation directly from the critical point to investigate the subsequent dynamics. Our previous numerical simulations have already investigated the catastrophe process, in which the slow evolution of the photospheric magnetic field drives the flux rope through a sequence of quasi-static states toward the loss of equilibrium at a critical point \citep[e.g.,][]{2012SCPMA..55.1316M,2017AcASn..58...55X,2022ApJ...933..148C,2024ApJ...962...42H}. These numerical experiments confirm that the system generally evolves along the path predicted by the catastrophe theory and eventually erupt nearby the critical points. In this context, the MHD simulations in this work are still performed following the catastrophe theory for CME eruptions.

\section{Results} \label{sec:result}

\subsection{Analytic Implications} \label{sec:result_analytic}

While the analytic model does not well resolve the reconnection related dynamics after the loss of equilibrium, it provides clear and testable guidance for the subsequent evolution and its underlying mechanisms. Figure~\ref{fig:curve} motivates three questions that frame our numerical study.

First, can the NEF drive the system toward catastrophe? Figure \ref{fig:curve} shows that catastrophe is attainable for both $S<0$ or $S>0$ as displayed by purple and orange points, implying that an NEF with either polarity can trigger the instability and initiate a filament eruption. An important exception occurs when a positive-polarity NEF emerges very close to the flux rope, which is excluded from our parameter survey and not shown in Figure \ref{fig:curve}.
 
Second, how does the system evolve once catastrophe occurs? For $S>0$, the equilibrium topology in Figure~\ref{fig:curve} varies with $x_d$. At $x_d=4$, the vertical dashed line $S=S^*$ intersects the equilibrium branches twice (orange point and star), indicating the existence of a higher equilibrium. At $x_d=7$, only one intersection remains, i.e., no higher equilibrium is available. The intermediate case $x_d=4.704$ is tangential, marking the boundary between these two topologies. In contrast, for $S<0$, once the system reaches the critical state (purple points), no further equilibrium exists. Taken together, these analytic results suggest the following expectations for the dynamics: the erupted flux rope will rise continuously to develop into a successful CME if no higher equilibrium position exists. On the other hand, if a higher equilibrium position exists, the flux rope is expected to jump to that position, representing a failed eruption. However, the meaning of the special intermediate state at $x_d=4.704$ for the flux rope dynamics remains unclear. In the following simulations, we will demonstrate that it in fact corresponds to more complex behavior, such as two-step eruptions.

Third, what are the asymmetric characteristics of the magnetic configuration at the onset of catastrophe, and how do they influence CME deflection? The asymmetry of the critical state, primarily quantified by $|S^*|$, sets the lateral Lorentz–force imbalance that drives non-radial motion. In Figure~\ref{fig:curve}, $|S^*|$ increases with $x_d$ for $S<0$ (more asymmetric critical states), whereas it decreases with $x_d$ for $S>0$ (more symmetric critical states; see the purple/orange markers and the corresponding field configurations). Because $|S^*|$ implies the departure from symmetry, we anticipate a monotonic relation between $|S^*|$ and the deflection angle $|\theta|$ (see Figure \ref{fig:config}) observed during the dynamic phase.

These analytic results provide crucial insights into the role of NEF in determining eruption outcomes. In the following section, we will perform resistive MHD simulations that include more realistic magnetic energy release and transfer processes (e.g., magnetic reconnection) to verify and refine these theoretical predictions.

\subsection{Numerical Results} \label{sec:result_numerical}

\subsubsection{Loss of equilibrium from critical states}\label{sec:loss}

\begin{figure}
\begin{interactive}{animation}{3.evolution_Jz_movie.mp4}
\includegraphics[width=0.95\textwidth]{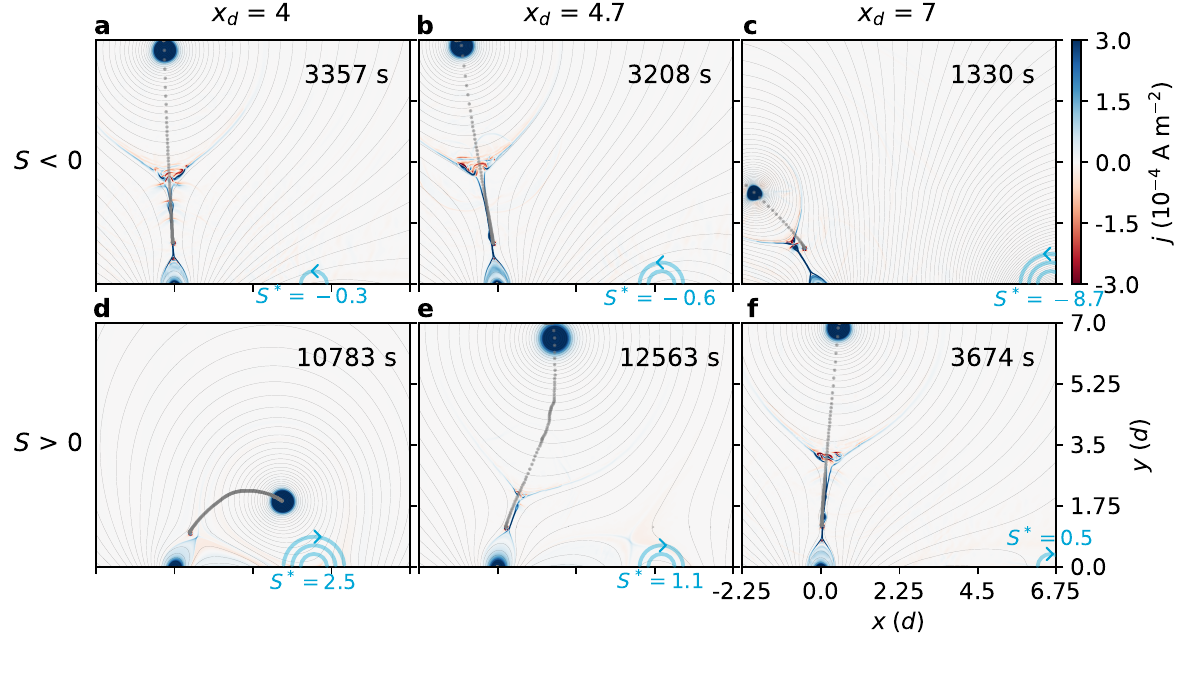}
\end{interactive}
\caption{Current density evolution in the corona, showing the dynamics of filament eruptions. Top and bottom rows correspond to $S<0$ and $S>0$, respectively; from left to right the columns are $x_d=4$, $4.7$, and $7$. Times in each panel are are measured from the initial equilibrium to the state shown. Background colors show the out-of-plane current density, and black curves are magnetic field lines. Cyan rings indicate the NEF, highlighting its polarity, horizontal location $x_d$, and the critical strength $S^*$. Overplotted gray dots trace the time history of the flux-rope center. This figure is available as an animation. Each animation presents the full temporal evolution for its corresponding case, covering physical times from $1.33\times10^{3}$ to $1.34\times10^{4}$~s, while the static panels show the final frames.
\label{fig:evo}}
\end{figure}

To display the dynamics when the predicted catastrophe is triggered, Figure~\ref{fig:evo} shows evolutions of current density in six representative runs, each initialized from the corresponding critical state in Figure~\ref{fig:curve}. Background shading shows current density, black curves denote magnetic field lines, cyan rings illustrate the NEF, and dots trace the flux-rope center. The full set of flux-rope trajectories across the parameter space is presented later (Figure~\ref{fig:traj}).

Starting from the critical states given by theory, the magnetic configuration in all cases loses equilibrium: the flux rope rises, a long current sheet forms between the flux rope and the closed magnetic loops (or flare loops) above the photosphere, the magnetic reconnection occurs associated with plasmoid formation, and the system transitions into a dynamic phase. Thus, within the surveyed parameter space, simulations confirm that the catastrophe indicated by the theory will lead to the onset of filament eruption.

Figure \ref{fig:evo} shows that an eruptive flux rope does not always develop directly into a CME; instead, it exhibits diverse kinematics, depending on the NEF parameters. In addition, the trajectories of flux rope show the non-radial propagation. The following subsections quantify these eruption modes and the associated deflection.

\subsubsection{Three eruption modes: one-step, two-step, and failed}\label{sec:modes}

\begin{figure}[ht]
\begin{center}
\includegraphics[width=0.65\textwidth]{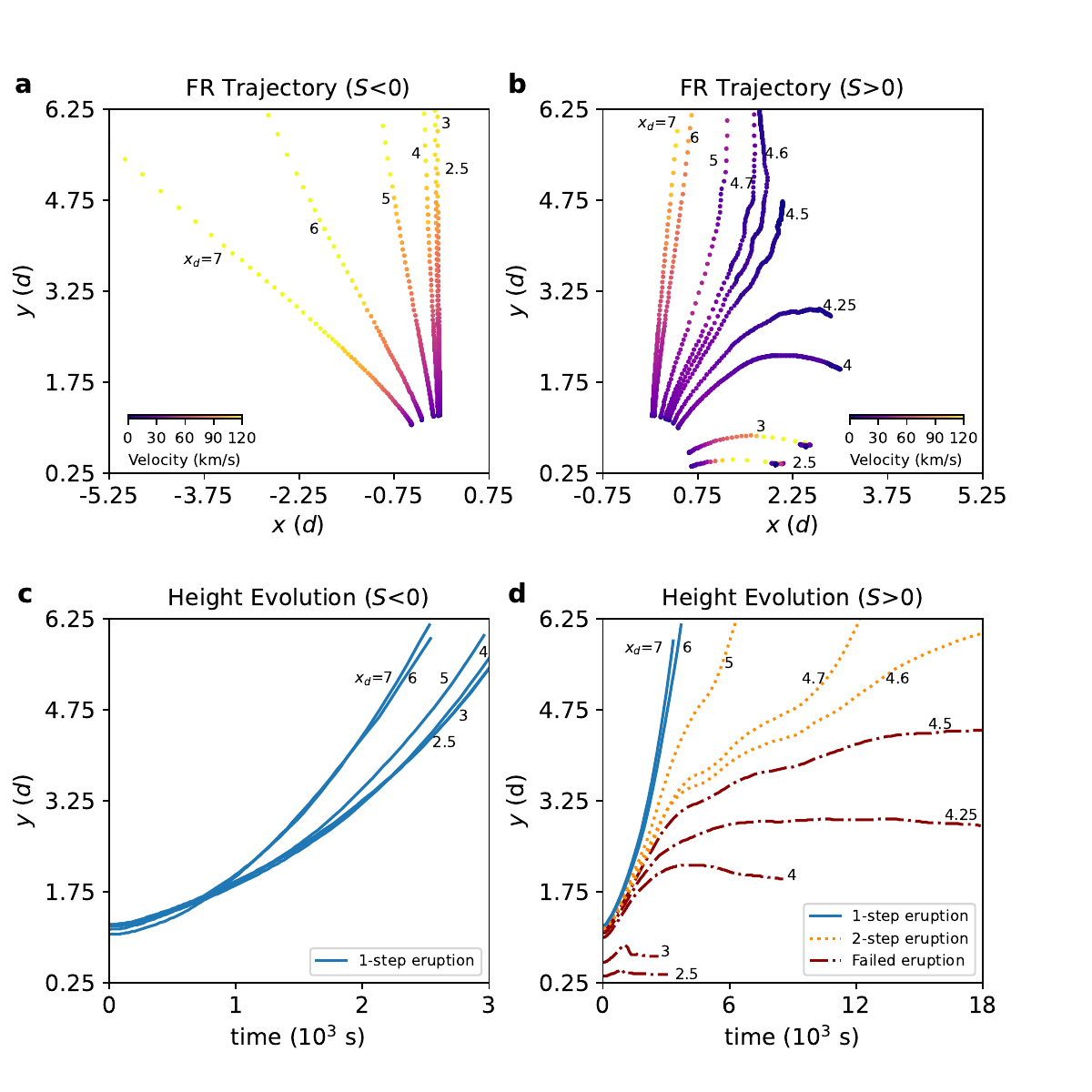}
\end{center}
\caption{The trajectories of flux-rope center. Top panels show the spatial $x$-$y$ path, while bottom panels present the temporal $y$-$t$ evolutions. The left and right columns correspond to NEF with $S<0$ and $S>0$, respectively. Different locations ($x_d$) are texted to corresponding results. Blue, orange, and red profiles at the bottom represent one-step, two-step, and failed eruptions, respectively.
\label{fig:traj}}
\end{figure}

Figure~\ref{fig:traj} compiles trajectories of the flux-rope center across the full parameter survey: the top and bottom panels plot the $x$-$y$ paths and the $y$-$t$ profiles, while the left and right columns correspond to cases with $S<0$ and $S>0$, respectively. Based on these diagnostics, we identify three representative eruption modes.

One-step eruptions: After the loss of equilibrium, the flux rope accelerates continuously and rapidly develops into a CME. As shown in Figure~\ref{fig:traj}, all $S<0$ (see panels a and c) simulations belong to this class; for $S>0$ (see panels b and d), one-step behavior also appears at larger horizontal positions (e.g., $x_d=6$ or $7$). Figure \ref{fig:traj}a suggests that speeds of one-step eruption typically reaches 100~km~s$^{-1}$ well below $2\times10^8$~m, which will likely lead to the CME as the flux rope propagates to the higher corona.

Two-step eruptions: After an initial rapid rise, the flux rope decelerates and exhibits a short plateau (sometimes with small oscillations), then re-accelerates to form a CME. In our survey this mode occurs for $S>0$ at $x_d=4.6$, $4.7$, and $5.0$ (as shown by orange dotted lines in Figure \ref{fig:traj}d). Owing to the mid-course slowdown, the speed at $2\times10^8$~m is markedly lower than in one-step cases, about 30~km~s$^{-1}$. The plateau lasts $0.5$-$2$~h in various runs, comparable in order of magnitude to hour-scale pauses in observations \citep{2014SoPh..289.4545B,2016ApJ...821...85G,2018MNRAS.475.1646F}, though much shorter than the rare $>10$~h events \citep{2017SoPh..292...81C}.

Failed eruptions: The flux rope rises after catastrophe but ultimately settles into a new equilibrium and does not evolve into a CME. This mode occurs for $S>0$ with $x_d\le4.5$ (Figure \ref{fig:traj}). As shown in Figure \ref{fig:traj}b, the kinematics for failed eruptions fall into three categories: slow rise followed by stalling ($x_d=4.5$); rise then fall toward a new equilibrium ($x_d=4.25$, $4$); and nearly horizontal motion with speed up to 100~km~s$^{-1}$ ($x_d=3$, $2.5$), similar to the transverse eruptions reported in observations \citep{2015SoPh..290.1703M}. These observed transverse eruptions, however, may result from multiple physical mechanisms, whereas those reproduced in our simulations are driven exclusively by the NEF. Therefore, not all observed transverse eruptions necessarily require an NEF with $S>0$ and low $x_d$. In future work, we plan to examine specific transverse eruptions associated with flux emergence and test whether their deflection angles and background magnetic configurations are consistent with our results.

The three modes seen in Figures \ref{fig:evo} and \ref{fig:traj} are in close agreement with the analytic guidance in Figure \ref{fig:curve}, which essentially depend on whether a new equilibrium exists along $S=S^*$. According to \cite{2001JGR...10625053L}, the existence of such an equilibrium requires that the magnetic energy $E_h(x_h, y_h)$ satisfies Equations \ref{Global eqx} and \ref{Global eqy}, usually corresponding to a local minimum of $E_h$.

If no higher equilibrium position exists beyond the critical point, $E_h$ decreases monotonically with increasing $y_h$, so the Lorentz force is insufficient to produce a downward restoring component. As a result, the erupting flux rope remains unstable throughout its low-coronal propagation, and its acceleration continues. Such unstable magnetic configurations at higher corona can be generated by two types of NEFs. In cases with $S<0$ at smaller $x_d$ (Figure \ref{fig:evo}a) or $S>0$ at larger $x_d$ (Figure \ref{fig:evo}f), the NEF is more likely to reconnect with the background field \citep{2000ApJ...545..524C}, thereby reducing the downward tension and preventing a new equilibrium from forming. In cases with $S<0$ at larger $x_d$ (Figure \ref{fig:evo}c), although the NEF is classified as reconnection-unfavorable \citep{2000ApJ...545..524C}, it primarily increases the magnetic pressure lateral to the flux rope, which amplifies the sideward Lorentz force and drives instability in both the vertical and horizontal directions.

If a new equilibrium exists, a local minimum of $E_h$ forms in higher corona, so the ascending flux rope will slow down and even be trapped in the minimum $E_h$ location. The new equilibrium can be produced by an NEF with $S>0$ at smaller $x_d$. As shown by the orange configuration in Figure~\ref{fig:curve}d, the emerging fields close to the filament (marked as orange ``$-$'') leads to the reconnection with the inner background field (black ``$+$'') and initiates the local instability, while the fields on the other side (orange ``$+$'') preferentially connect to outer background fields overlying the flux rope (black``$-$''), continually enhance the global stability. The extra flux over the flux rope creates a new stable zone in the higher corona. This explains why failed eruptions appear only for $S>0$ in our runs (see Figure~\ref{fig:traj}). 

If a new equilibrium just emerges, the two equilibrium branches become tangent (see Figure~\ref{fig:curve}, $S>0$ with $x_d=4.704$). This special equilibrium additionally satisfies the critical point condition (Equation \ref{critical point}) and should therefore be regarded as an inflection point of $E_h$ governed by multiple parameters \citep{2001JGR...10625053L}. In other words, this equilibrium is only metastable, and the analytic model alone cannot determine the outcome. Our simulations show that the flux rope pauses briefly near this metastable position and then loses equilibrium again, producing a two-step eruption (Figure~\ref{fig:evo}e). 

These results indicate that the NEF plays a key role in reshaping coronal stability and thereby influences the kinematics of eruptive flux ropes.

\subsubsection{Non-radial propagation of eruptive filaments deflected by the NEF}\label{sec:deflection}

\begin{figure}[ht]
\begin{center}
\includegraphics[width=0.65\textwidth]{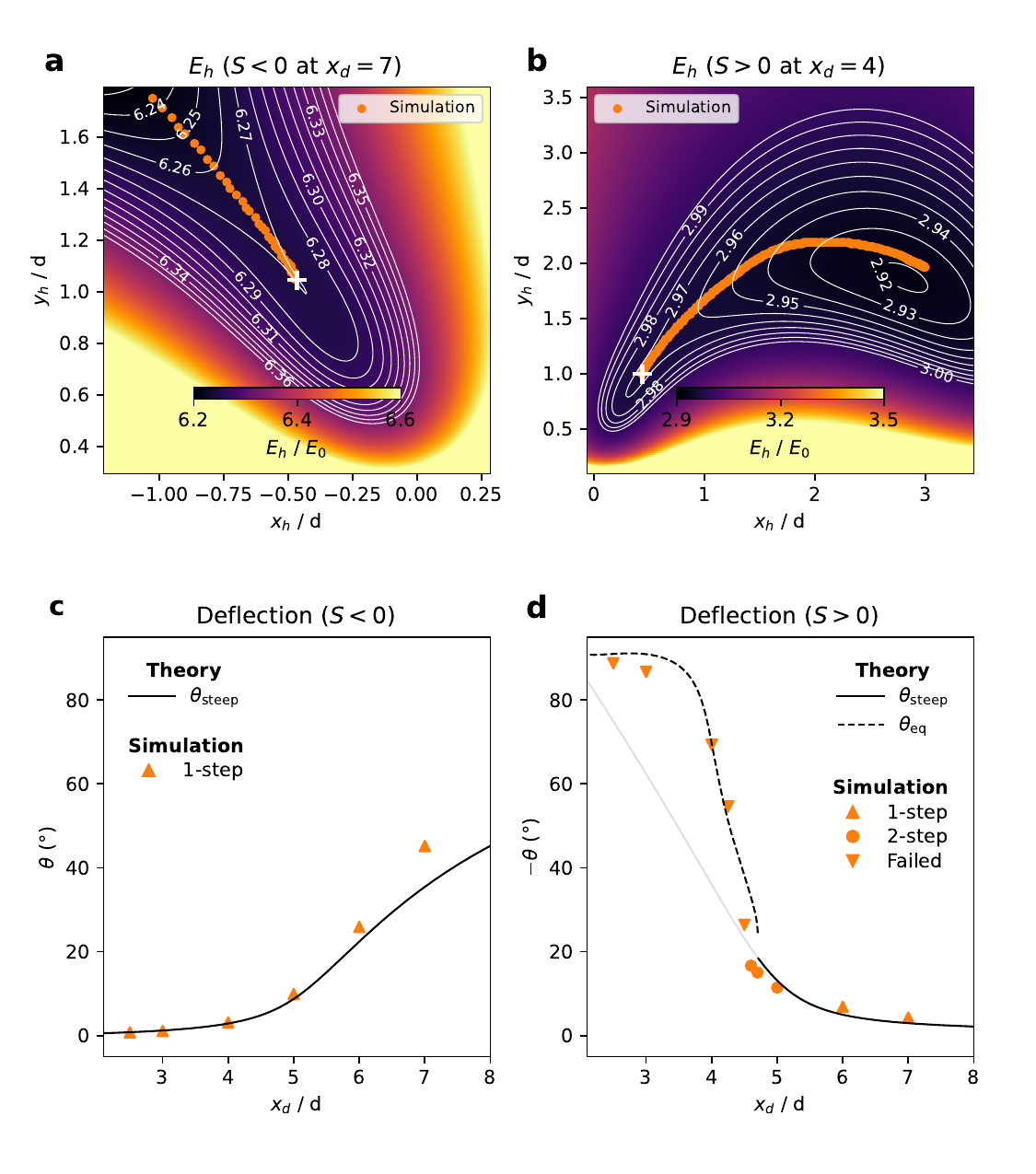}
\end{center}
\caption{Top panels show the magnetic energy $E_h$ (Equation \ref{Eh}) of the system as a function of the flux-rope position $(x_h,y_h)$ around the critical points: (a) a succesful eruption (or CME) with $S<0$ and $x_d=7$; (b) a failed eruption with $S>0$ and $x_d=4$. Background shading and white contours show normalized $E_h$, the white ``+'' marks the critical equilibrium, and orange dots trace the simulated trajectory of the FR center. Bottom panels display the deflection angle $\theta$ versus $x_d$ for these two cases: (c) $S<0$ and (d) $S>0$. The black curves are theoretical angle profiles: the solid curve is $\theta_{\mathrm{steep}}$ (steepest-decrease direction of $E_h$ at the critical point), and the dashed curve is $\theta_{\mathrm{eq}}$ (direction from the critical point to the secondary equilibrium). Orange symbols denote eruption modes from simulations: up-pointing triangles (one-step), circles (two-step), and down-pointing triangles (failed). In panel (d), we plot $-\theta$ for sign consistency. Overall, simulated CME deflections follow $\theta_{\mathrm{steep}}$, whereas failed eruptions align with $\theta_{\mathrm{eq}}$.
\label{fig:deflection}}
\end{figure}

When the eruption develops into a CME, especially for one-step events, the trajectory is nearly rectilinear (Figure \ref{fig:traj}a), suggesting that the CME direction can be inferred from the most favorable escape direction near the critical state. To find this direction, we compute the distribution of the magnetic energy $E_h(x_h,y_h)$ around the critical equilibrium from Equation~\ref{Eh}, and display a representative CME case, $S<0$ with $x_d=7$, in Figure~\ref{fig:deflection}(a). Shading and contours show normalized $E_h$, with the characteristic energy $E_0=(I_0/cd)^2$. The white “+” marks the critical equilibrium, and orange dots trace the path from the above resistive MHD simulation. 

Figure \ref{fig:deflection}(a) clearly suggests that the trajectory of CMEs aligns with the direction along which $E_h$ decreases most rapidly. We denote this direction by $\theta_{\mathrm{steep}}$ (the definition of $\theta$ follows Figure~\ref{fig:config}). Near the critical point, most directions show positive energy increments and a restoring tendency of the flux rope, whereas along $\theta_{\mathrm{steep}}$ the energy increment vanishes and the flux rope is likely to escape. In general, $\theta_{\mathrm{steep}}$ aligns with the negative gradient of $E_h$. At the critical point $(x_h^{*},y_h^{*})$, however, the first derivatives vanish because of the global equilibrium (Equations~\ref{Global eqx} and \ref{Global eqy}), while the second derivatives satisfy the critical–point condition
(Equation~\ref{critical point}). By further calculation, we find the $\theta_{\mathrm{steep}}$ can be obtained with the second derivatives of $E_h$ by:
\begin{equation}
\theta_{\mathrm{steep}}=\arctan\!\frac{\partial^2 E_h/\partial x_h\partial y_h}{\partial^2 E_h/\partial x_h^2},
\label{def_angle}
\end{equation}
where the detailed expression for these derivatives are displayed in Equations \ref{Exx}-\ref{Exy}. See more details in Appendix~\ref{sec:appA}.

For failed eruptions (e.g., $S>0$ and $x_d=4$ in Figure~\ref{fig:deflection}b), a minimum of $E_h$ forms in higher corona, and the trajectory departures from a straight line, so $\theta_{\mathrm{steep}}$ only shows the early tendency and no longer reproduces their eventual directions. We therefore introduce an alternative predictor from the theory, $\theta_{\mathrm{eq}}$, defined as the angle between the critical state $(x_h^*,y_h^*)$ and the secondary equilibrium $(x_h^{\mathrm{2nd}},y_h^{\mathrm{2nd}})$ at the same $S^*$ (see the orange dot and star in Figure~\ref{fig:curve}):
\begin{equation}
\theta_{\mathrm{eq}}=-\arctan\!\frac{x_h^{\mathrm{2nd}}-x_h^*}{\,y_h^{\mathrm{2nd}}-y_h^*\,}.
\label{def_angle2}
\end{equation}
The critical state $(x_h^*,y_h^*)$ follows from Equation~\ref{critical point}, while $(x_h^{\mathrm{2nd}},y_h^{\mathrm{2nd}})$ corresponds to the other branch of the equilibrium at the same $S^*$ (Equations~\ref{Global eqx}--\ref{Global eqy}). Although explicit expressions are unavailable, these quantities can be readily obtained by numerical computation \citep{2024ApJ...977L..26C}.

Figures~\ref{fig:deflection}(c) and \ref{fig:deflection}(d) compile the theoretical profiles of $\theta_{\mathrm{steep}}$ (solid) and $\theta_{\mathrm{eq}}$ (dashed) versus $x_d$ for $S<0$ and $S>0$, respectively, with simulations overplotted as up-pointing triangles (one-step), circles (two-step), and down-pointing triangles (failed). For CME cases, $\theta$ is measured when the flux-rope center reaches $2.5\times10^8$~m; for failed cases, $\theta$ is measured at the final settling time. For display consistency, panel~(d) shows $-\theta$.

Figure~\ref{fig:deflection} suggests that the NEF polarity determines the sign of the lateral deflection. For $S<0$, the NEF repels the flux rope, driving a horizontal drift away from the NEF ($\theta>0$); for $S>0$, it attracts the flux rope, yielding a drift toward the NEF ($\theta<0$). According to \cite{2024ApJ...977L..26C}, the horizontal component of the Lorentz force exerted by the NEF on the flux rope aligns with the sign of $S$. We further point out a nearby NEF with $S<0$ ($S>0$) will increase (decrease) the local magnetic energy, thus repels (attracts) the flux rope.

As established in previous work, coronal deflectors are often grouped into two phenomenological families: repulsors and attractors (see reviews in \citealt{2023FrASS..1060432C}). This practice usually assigns each structure to a single category—for example, coronal holes are typically labeled as repulsors, whereas the heliospheric current sheet is labeled as an attractor. Our results indicate that this assumption is not generally valid. The NEF can act as a repulsor or an attractor depending on its polarity relative to the background field. In addition, other simulations have shown that a coronal hole of opposite polarity can first attract and then repel a CME \citep{2021A&A...652A.111S}. Taken together, these behaviors show that CME deflection by coronal magnetic structures cannot be fully captured by a simple attractor-repulsor classification.

Having analyzed the sign of $\theta$, we now quantify the magnitude of the deflection and its physical origin. Figure~\ref{fig:deflection} shows that the deflection angle depends on both $S$ and $x_d$: for $S<0$, $|\theta|$ increases with $x_d$, whereas for $S>0$, $|\theta|$ decreases with $x_d$. This trend tracks the analytic behavior of the critical strength $|S^*|$ (Figure~\ref{fig:curve}; see also Section~\ref{sec:result_analytic}). Because $|S^*|$ measures how strongly the NEF breaks the symmetry of the overlying field, larger $|S^*|$ implies a larger lateral imbalance and hence a larger $|\theta|$, while smaller $|S^*|$ corresponds to a more symmetric configuration and a smaller $|\theta|$.

To interpret this dependence, \citet{2024ApJ...977L..26C} introduced a filament-channel function,
\begin{equation}
C(x_d, y_d)=x_d-\sqrt{\frac{y_d+3+2\ln\!\left(2/r_{00}\right)}{-y_d+1+2\ln\!\left(2/r_{00}\right)}}\,(y_d+1).
\label{channel}
\end{equation}
for which $C<0$, $C=0$, and $C>0$ locate the NEF inside, at the boundary of, and outside the filament channel, respectively (for parameters used here, this corresponds to $x_d<5.2$, $x_d=5.2$, and $x_d>5.2$). Following \citet{2024ApJ...977L..26C}, an NEF inside the channel with $S<0$ or outside the channel with $S>0$ tends to trigger the eruption, indicating a weaker critical strength $|S^*|<1$; conversely, an NEF inside with $S>0$ or outside with $S<0$ requires stronger emergence to erupt, implying $|S^*|>1$.

Therefore, the product $S\!\cdot\!C$ not only diagnoses whether the NEF is likely to trigger eruptions, but also describe the symmetry when eruptions onset. When $S\!\cdot\!C>0$ (e.g., $S<0$ with $x_d<5.2$; $S>0$ with $x_d>5.2$ in Figure~\ref{fig:deflection}), the coronal structure remains more symmetric ($|S^*|<1$), and the deflection stays small ($|\theta|<11^\circ$ in our simulations). Conversely, when $S\!\cdot\!C<0$ (e.g., $S<0$ with $x_d>5.2$; $S>0$ with $x_d<5.2$), the configuration becomes more asymmetric ($|S^*|>1$), and the deflection becomes larger ($|\theta|> 11^\circ$). Within our explored parameter space, $|\theta|$ can reach $\sim45^\circ$ for $S<0$ and approach $\sim90^\circ$ for $S>0$. 

Figures~\ref{fig:deflection}(c) and \ref{fig:deflection}(d) show that, overall, the theoretical predictors $\theta_{\mathrm{steep}}$ and $\theta_{\mathrm{eq}}$ reproduce the simulated deflections of CMEs and failed eruptions, particularly at small $|\theta|$. At larger $|\theta|$, systematic departures appear. For $S<0$, for example at $x_d=7$, the simulated deflection angles tend to exceed the analytic values. A qualitative explanation is that the flux-rope current $J$ decreases during the eruption as the reconnection inside it occurs, whereas the NEF field near the flux rope changes little. As the NEF field becomes relatively stronger, it amplifies the lateral force and the simulated deflection exceeds the analytic estimate. For $S>0$, for example at $x_d=3$, the simulated deflections are slightly smaller than predictions because reconnection between the NEF and the background fields weakens the NEF field and reduces the lateral force.

These results indicate that the NEF shapes non-radial propagation by fixing the sign of the deflection through its polarity and by setting the magnitude through the asymmetry it introduces.

\section{Discussions} \label{sec:discussions}

While \citet{2024ApJ...977L..26C} demonstrated how an NEF can drive a filament to loss of equilibrium, they could not address the subsequent dynamics due to their analytic framework. Based on their theory, we perform resistive MHD simulations to examine how the NEF conducts the consequent propagation of eruptive filaments in the low corona. The simulations reveal diverse outcomes of eruptive filaments, CMEs with ``one-step'' or ``two-step'' evolution, and failed eruptions (Section~\ref{sec:modes}), as well as pronounced non-radial deflection. These behaviors arise because the NEF reshapes the magnetic configuration and introduces asymmetry into the coronal field. Unlike most prior work that focuses only on triggering eruptions, we highlight the NEF’s impact on the early propagation of solar eruptions.

Classical catastrophe theory emphasizes vertical loss of equilibrium in symmetric backgrounds and neglects horizontal instabilities \citep{1991ApJ...373..294F,2005IAUS..226..250R}. In the asymmetric corona, the loss-of-equilibrium direction is no longer radial, so the preferred escape direction must be identified. \cite{2001JGR...10625053L} constructed the ideal MHD model involving the asymmetric configuration, but they do not address this direction. Comparing theory with simulations, we find that the direction of the steepest decrease of the magnetic energy at the critical point, $\theta_{\mathrm{steep}}$ (Equation~\ref{def_angle}), predicts the early escape direction of filaments and thus the CME trajectory in the low corona. At greater heights and into interplanetary space, the ambient solar wind is expected to dominate the propagation. In future work we will embed these onset states in large-scale, global MHD simulations including solar wind to quantify how the early direction affects the eventual CME direction.

Previous numerical studies have shown that the presence (absence) of a higher equilibrium predicted by the catastrophe theory corresponds to a failed eruption (CME) \citep{2022ApJ...933..148C,2024ApJ...977L..26C}. Here we notice a transitional case in which the two equilibrium branches become tangent (Figures~\ref{fig:curve}b and \ref{fig:curve}e), signaling that a higher equilibrium has just appeared. The dynamics initiated from this special case was unknown. Our simulations show that this state is metastable and produces a ``two-step'' evolution. The mechanism of forming a higher metastable equilibrium is akin to \citet{2018MNRAS.475.1646F}, who realized ``two-step'' behavior in a symmetric background created by two dipoles located at the same horizontal position; in contrast, our configuration is intrinsically asymmetric due to the NEF, providing a new explanation for two-step eruptions.

The influence of NEF on the subsequent propagation of CMEs has received far less attention than its triggering role. With improving observations and advancing simulations this topic is gaining momentum—for example, \citet{2025ApJ...988L..36I} report that a very small NEF barely affects the onset of eruption but apparently accelerate the main phase. Here we systematically show that the NEF reshapes the coronal environment and thereby governs dynamics. First, it reshapes stability in the low corona, modulating various acceleration phases. Second, it breaks the symmetry in coronal configurations and deflects the eruption toward the side with lower magnetic energy, yielding non-radial eruption. Recent observations show that the asymmetric poloidal component of the coronal field influences both acceleration and direction \citep{Qiu2025KinematicalFilamentsPoloidalB}, which are consistent with this picture. 

Magnetic reconnection plays an essential role in producing the eventual eruption. The ideal loss of equilibrium initiates the motion, but the subsequent evolution requires reconnection to modify the magnetic topology and release magnetic energy \citep{2003NewAR..47...53L}. In the asymmetric configurations imposed by the NEF, the corona hosts one or two current sheets that control the kinematics. (i) When no higher equilibrium exists, a long current sheet forms below the flux rope; reconnection there produces concave-up reconnected arcs whose upward tension accelerates the rope, yielding a one-step eruption \citep{2021NatAs...5.1126J}. (ii) When a higher equilibrium exists, the NEF compresses the magnetic field of the flux rope, generates a current sheet between them, and drives the magnetic reconnection \citep[e.g., see][]{2025SCPMA..6895211X}; reconnection in this sheet threads additional overlying flux and can halt the ascent, leading to failed eruptions \citep{2022ApJ...933..148C}. 
(iii) For cases with a metastable equilibrium, the acceleration and deceleration induced by reconnection in the two current sheets partly balance, leading to a two-step behavior. If $4.6<x_d<4.704$, although a higher equilibrium exists in the ideal analysis, fast reconnection in the sheet below the flux rope ultimately produces the eruption. If $4.704<x_d<5.0$, the analytic model predicts no higher equilibrium position, so the system naturally evolves into a CME as expected. In this case, reconnection in the current sheet between the flux rope and the NEF temporarily delays the acceleration and produces the two-step eruption. This explains why the simulations exhibit two-step eruptions over the finite interval $4.6<x_d<5.0$, rather than only at the single tangency $x_d=4.704$ predicted by the theory.

Although our 2D model simplifies the magnetic configuration, it captures the basic physics needed to interpret CME deflections in observations. In particular, for eruptions driven by an NEF that can be approximated as a bipole near a filament channel, the model yields a semiquantitative diagnostic for deflection forecasting. First, measure the sign of $S$ and the horizontal offset $x_d$ directly from the photospheric magnetogram, and infer the NEF magnitude $|S|$ and depth $y_d$ following the method in \citet{2001JGR...10625053L}. Second, compute the filament-channel function $C(x_d,y_d)$ (Equation~\ref{channel}). Then $S\!\cdot\!C>0$ indicates that the eruption is expected to be nearly radial (small $|\theta|$), whereas $S\!\cdot\!C<0$ implies that the strong non-radial propagation is likely (large $|\theta|$). For more general coronal magnetic topologies where the filament-channel scenario does not apply, we estimate $|\theta|$ using Equation~\ref{def_angle}, which depends on the gradient of the magnetic energy at the critical point and does not assume a specific topology. This approach requires coronal magnetic-field measurements, a challenging area with notable recent progress \citep[e.g.,][]{2020Sci...369..694Y,2020ScChE..63.2357Y,2024Sci...386...76Y}. Multi-height magnetic-field observations across the solar atmosphere will be essential to validate and calibrate these diagnostics.

Furthermore, our model sheds light on several questions associated with non-radial eruptions. First, about 75\% of events are nearly radial \citep{2015SoPh..290.1703M}, despite ubiquitous asymmetry in the photospheric magnetic flux. One explanation is that stronger asymmetry in the coronal field reduces non-potential magnetic energy that mainly powers eruptions \citep{2024MNRAS.533L..25L}. Our results add a complementary view: large inclinations generally require an NEF much stronger than the background field ($|S^*|\!\gg\!1$), which is less frequent than cases with $|S|<1$ and $|S|\approx1$, so strongly oblique events should be intrinsically rare. Second, the lateral expansion of a CME launches a coronal EUV wave whose chromospheric footprint is the Moreton wave \citep{1968SoPh....4...30U}. Yet most EUV waves are not accompanied by a Moreton counterpart \citep{2025ApJ...980..254W}. A likely explanation is that Moreton waves occur preferentially during sufficiently inclined eruptions \citep{2023ApJ...949L...8Z,huang20253dfastmodewavepropagation}. The prediction of non-radial CMEs may help us identify Moreton waves in the early stage. Additionally, in some inclined eruption cases, multiple EUV waves have been observed without accompanying Moreton waves. These studies (e.g., \citealt{2024SCPMA..6759611Z,2024ApJ...974L...3Z}) found that the formation of these wavefronts is not only closely related to the inclination angle of the eruption filaments, but also shows a significant association with the unwinding of the filaments. Quantifying how the deflection angle maps to the direction and amplitude of the Moreton signature warrants further study. Finally, the diagnostics for non-radial eruptions provide useful constraints for 3D reconstructions of inclined CMEs \citep{2021A&A...653L...2Z,2025ApJ...985..237Z} and tilted trailing current sheets \citep{2023ApJS..269...22C}.

\section{Conclusions}
\label{sec:conclusion}

In this work, we performed 2D MHD simulations to investigate how an NEF with polarity $S$ and horizontal offset $x_d$ drives non-radial eruptions. Our results suggest that the NEF not only triggers filament eruptions but also organizes the subsequent kinematics by reshaping coronal asymmetry. Our main conclusions are:

(1) The development of the flux rope, initialized from various critical points of the catastrophe theory, is revealed by using resistive MHD simulations. Simulations show that filaments erupt and enter the dynamic phase, confirming the predicted loss of equilibrium.

(2) The acceleration phases of the flux rope depend on $S$ and $x_d$. For $S < 0$ (the polarity of NEF is negative compared with the original background sources), the flux rope continuously accelerates and develops to a CME (one-step). For $S>0$ (the polarity of NEF is positive), the outcome varies with $x_d$: when the NEF is far from the flux rope ($x_d>5$), the eruption is one-step; when it is close ($2.5<x_d<4.6$), the flux rope stops at a new equilibrium and the eruption fails; at an intermediate position ($4.6\le x_d\le 5$), the rising rope decelerates and stalls at a higher height, then re-accelerates to form a CME (two-step). These diverse acceleration phases arise because the NEF redistributes the equilibrium in corona. When eruptions onset: if no higher equilibrium exists, the magnetic energy $E_h$ decreases monotonically with height, and one-step eruption is obtained; if a higher equilibrium exists, $E_h$ attains a minimum, and the failed eruption occurs; if the high equilibrium just emerged, $E_h$ reaches the metastable inflection point, and the two-step eruption is likely to occur.

(3) By introducing asymmetry into the corona, the NEF deflects solar eruptions from the radial direction. For $S<0$ the NEF repels the flux rope; for $S>0$ it attracts the flux rope. The deflection amplitude also depends on $x_d$: for $S<0$ ($S>0$), $|\theta|$ increases (decreases) with $x_d$, reflecting the growing (diminishing) asymmetry in the magnetic configuration. We derive two characteristic angles from the theory to predict deflection, and they can be evaluated using only the local magnetic configuration at the critical point. $\theta_{\mathrm{steep}}$, defined as the steepest-descent direction of $E_h$ at the critical point, represents the CME propagation direction; $\theta_{\mathrm{eq}}$, defined as the geometric angle between the critical position where the eruption begins and the higher equilibrium where the filament halts, characterizes the direction of failed eruptions. Both angles can be determined at the onset of solar eruptions, making them practical predictors of deflections in the low corona.




\vspace{\baselineskip} 
\noindent We gratefully acknowledge constructive comments given by Drs. Xiaoyan Xie, Ivan Oparin, Xinping Zhou, and John C. Raymond. This work was supported by the Strategic Priority Research Program of the Chinese Academy of Sciences No.XDB0560000, Specialized Research Fund for State Key Laboratory of Solar Activity and Space Weather, National Key R\&D Program of China No. 2022YFF0503800, NSFC grants 12573062, 12273107, and 12503064, grants associated with the Yunnan Revitalization Talent Support Program, the Foundation of the Chinese Academy of Sciences (Light of West China Program), the Yunling Scholar Project of Yunnan Province and the Yunnan Province Scientist Workshop of Solar Physics, and grants 202101AT070018 and 2019FB005 associated with the Applied Basic Research of Yunnan Province. C.Y. acknowledges the support from the University of Chinese Academy of Sciences Joint Ph.D. Training Program. The numerical computation in this paper was carried out on the computing facilities of the Computational Solar Physics Laboratory of Yunnan Observatories (CoSPLYO).



\appendix

\section{Derivation details for $\theta_{\mathrm{steep}}$}\label{sec:appA}

We consider a small displacement from the critical position $(x_h^*,y_h^*)$ of the flux rope, written as $(\Delta x,\Delta y)=(-r\sin\theta,\,r\cos\theta)$. Expanding the magnetic energy $E_h$ to second order gives 
\begin{align} \Delta E \;=\;& \frac{\partial E_h}{\partial x_h}\Delta x + \frac{\partial E_h}{\partial y_h}\Delta y +\frac{1}{2}\!\left[ \frac{\partial^2 E_h}{\partial x_h^2} (\Delta x)^2 + 2\frac{\partial^2 E_h}{\partial x_h\partial y_h}\Delta x\Delta y + \frac{\partial^2 E_h}{\partial y_h^2} (\Delta y)^2 \right] + \cdots \notag \\ \approx\;& \frac{1}{2}\,r^2 \!\left( \frac{\partial^2 E_h}{\partial x_h^2}\,\sin^2\theta - 2\,\frac{\partial^2 E_h}{\partial x_h\partial y_h}\,\sin\theta\,\cos\theta + \frac{\partial^2 E_h}{\partial y_h^2}\,\cos^2\theta \right), \end{align} where first derivatives vanish (see Equations \ref{Global eqx} and \ref{Global eqy}). The departures from $(x_h^*,y_h^*)$ along most directions increase the system energy ($\Delta E>0$), implying a restoring tendency, whereas along the special direction that minimizes $\Delta E$ the second-order increase in energy is absent, so the flux rope is likely to escape.

Minimizing $\Delta E$ with respect to $\theta$ gives
\begin{equation}
\tan\!\big(2\theta_{\mathrm{steep}}\big)=\frac{2\,\partial^2 E_h/\partial x_h\partial y_h}{\;\;\partial^2 E_h/\partial x_h^2-\partial^2 E_h/\partial y_h^2}.
\end{equation}
Using the catastrophe condition in Equation (\ref{critical point}), this general form reduces to the compact expression used in the main text:
\begin{equation}
\theta_{\mathrm{steep}}=\arctan\!\frac{\partial^2 E_h/\partial x_h\partial y_h}{\partial^2 E_h/\partial x_h^2}.
\label{def_angle_App}
\end{equation}

The current inside the flux rope and its position satisfy the flux-frozen condition at the flux rope surface (see \citealt{2001JGR...10625053L}):
\begin{align}
J\ln\left(\frac{2y_hJ}{r_{00}}\right)
+\frac{2M\left(y_h+1\right)}{x_h^{2}+\left(y_h+1\right)^{2}}
+\frac{2S\left(y_h+y_d\right)}{\left(x_h-x_d\right)^{2}+\left(y_h+y_d\right)^{2}} 
&= \ln\left(\frac{2}{r_{00}}\right)+1. \label{forzenflux_eq} 
\end{align}
Combining Equations~\ref{critical point} and \ref{forzenflux_eq}, the second derivatives entering Equation~\ref{def_angle_App} are
\begin{align}
\frac{\partial^2 E_h}{\partial x_h^2}&= \left(\frac{I_0}{cd}\right)^2
\Bigg\{
\frac{8 J M \, \left[ (y_h+1)^2 - 3 x_h^2 \right](y_h+1)}{\left[x_h^2 + (y_h+1)^2\right]^3} 
+
\frac{8 J S \, \left[ (y_h + y_d)^2 - 3 (x_h - x_d)^2 \right](y_h + y_d)}
{\left[(x_h - x_d)^2 + (y_h + y_d)^2\right]^3}
\Bigg\}, \label{Exx}
\end{align}
\begin{align}
\frac{\partial^2 E_h}{\partial y_h^2} &= \left(\frac{I_0}{cd}\right)^2 \Bigg\{ 
\frac{8 J M \left[3 x_h^2 - (y_h+1)^2\right](y_h+1)}{\left[x_h^2 + (y_h+1)^2\right]^3} 
+
\frac{8 J S \left[3 (x_h - x_d)^2 - (y_h + y_d)^2\right](y_h + y_d)}
{\left[(x_h - x_d)^2 + (y_h + y_d)^2\right]^3}
 \notag \\
&\qquad\qquad + \frac{J^2}{2y_h^2} \frac{1}{\ln \left(\frac{2 y_h J}{r_{00}}\right) + 1}
+ \frac{J^2}{y_h^2} \Bigg\},\label{Eyy}
\end{align}
\begin{align}
\frac{\partial^2 E_h}{\partial x_h \partial y_h} &= \left(\frac{I_0}{cd}\right)^2 
\Bigg\{
\frac{8 J M \, \left[x_h^2 - 3(y_h+1)^2\right]x_h}{\left[x_h^2 + (y_h+1)^2\right]^3} 
+
\frac{8 J S \, \left[(x_h - x_d)^2 - 3 (y_h + y_d)^2 \right](x_h - x_d)}
{\left[(x_h - x_d)^2 + (y_h + y_d)^2\right]^3}
\Bigg\}.
\label{Exy}
\end{align}
Substituting these expressions into Equation~\ref{def_angle_App} yields $\theta_{\mathrm{steep}}$ as a function of $x_d$ (Figure~\ref{fig:deflection}).


\bibliography{cyh}{}
\bibliographystyle{aasjournal}


\end{CJK*}
\end{document}